\documentstyle [epsfig,11pt]{article}
\newcommand{\PRL}[3]{Phys.~Rev.~Lett.\ {\bf #1}, #2 (#3)}
\newcommand{\PRB}[3]{Phys.~Rev.~B {\bf #1}, #2 (#3)}
\newcommand{\JPI}[3]{J.~Phys.~I (France) {\bf #1}, #2 (#3)}
\newcommand{\JPA}[3]{J.~Phys.~A {\bf #1}, #2 (#3)}
\newcommand{\JPC}[3]{J.~Phys.~C {\bf #1}, #2 (#3)}
\newcommand{\EUL}[3]{Europhys.~Lett.\ {\bf #1}, #2 (#3)}
\newcommand{\EPJB}[3]{Eur.~Phys.~J.~B {\bf #1}, #2 (#3)}
\begin{document}

\title{Reparametrization invariance: a gauge-like symmetry
of ultrametrically organised states
}
\author{T.~Temesv\'ari\\Institute for Theoretical Physics,
E\"otv\"os University,\\H-1117 P\'azm\'any P\'eter s\'et\'any 1, Bld.~A,
Budapest, Hungary
\and    I.~Kondor\\Department of Physics of Complex Systems,
E\"otv\"os University,\\
 H-1117 P\'azm\'any P\'eter s\'et\'any 1, Bld.~A,
Budapest, Hungary
\and
        C.~De Dominicis\\Service de Physique Th\'eorique,
	CEA Saclay,\\F-91191 Gif sur Yvette, France}
\maketitle
\begin{abstract}
The reparametrization transformation between ultrametrically
organised states of replicated disordered systems is explicitly
defined. The invariance of the longitudinal free energy under
this transformation, i.e.\ reparametrization invariance, is shown
to be a direct consequence of the higher level symmetry of replica
equivalence. The double limit of infinite step replica symmetry
breaking and $n\to 0$ is needed to derive this {\em continuous}
gauge-like symmetry from the {\em discrete} permutation invariance
of the $n$ replicas. Goldstone's theorem and Ward identities can
be deduced from the disappearence of the second
(and higher order) variation of
the longitudinal free energy. We recall also how these and other exact
statements follow from permutation symmetry after introducing the
concept of "infinitesimal" permutations.

\end{abstract}

\section{Introduction}
Exact statements such as Goldstone's theorem \cite{Go,GSW} and
Ward identities \cite{Ward} proved very useful in the perturbative
analysis of ordinary statistical systems with a continuous
symmetry. For example, Goldstone's theorem ensures that the low
temperature phase of the O($m$) model is massless (the transverse
susceptibility is infinite), while the Ward identities, i.e.\ the 
exact relations between the different vertex functions (derivatives of
the Legendre-transformed free energy), help us to get rid of dangerous
infrared divergences, order by order in perturbation theory
\cite{BWW}.

Zero modes and massless phases are surprisingly frequent in quenched
random systems, too. The idea of the spin glass phase being marginal
occured very early to various authors \cite{TAP,BM1}
studying the mean field version of the Ising spin
glass, the Sherrington-Kirkpatrick (SK) model
\cite{SK},
without using replicas. After the solution of the SK model by
Parisi \cite{Pa}, the marginality of one of the most dangerous
eigenvalues across the whole spin glass phase was proven by two
different technics \cite{So,Go1}. In the truncated version of the SK
model, valid close to $T_c$, a whole band of zero and near-zero
eigenvalues were found \cite{DoKo}. The nature of the spin glass
phase in finite dimensional models with short range interaction
has been the subject of intensive debate, and the question has not
been settled until now. Nevertheless, both conflicting theories,
"droplet" picture or phenomenological scaling on one side
\cite{McMillan,droplet1,droplet2} and the ultrametrically organised
complex phase space structure of Parisi on the other
\cite{Pa,KoDo,Review}, predict a marginal spin glass phase with an
infinite spin glass susceptibility.

To give another example of a quenched disordered system with a
massless low temperature phase, we can mention the long-range
correlated random manifold problem in $d=D+N$ dimensions, where $D$ is
the intrinsic dimension of the manifold. For $N\to \infty $,
while $D<4$ is kept fixed, the zero momentum limit of a family of
eigenvalues of the mass operator goes to zero \cite{manifold}.
(In the notation of Ref.~\cite{manifold}, $\lambda_{\vec
p=0}(x;x,x)=0$, $0\leq x\leq 1$, where the momentum vector
$\vec p$ is $D$ dimensional.)

The idea that a continuous symmetry emerges in the Ising spin glass,
which is a discrete model, giving rise to Goldstone modes was raised
twenty years ago \cite{BM2}. Bray and Moore argued, that it is
the replica
limit $n\to 0$, which may be responsible for this rather strange
phenomenon. In Sec.~\ref{inf}, we will recall how the permutation
invariance of the $n$ replicas becomes a continuous symmetry
in systems whose ordered phase is organised in the hierarchical way
proposed by Parisi, but only in the limit $R\to \infty$, where
$R$ is the number of steps in Parisi's construction. All the
examples above belong to this class, except the "droplet" theory,
which corresponds to a replica symmetric ($R=0$) picture.
The symmetry argument we will use applies for a replicated system,
and is a direct consequence of the replica trick. How to derive
Goldstone's theorem in the original models and what is the
continuous symmetry there, is not clear for us and remains an open
question. The results of Sec.~\ref{inf} were
derived earlier using different, though closely related,
infinitesimal permutations \cite{Cyrano}.

The main result of this paper is left to Sec.~\ref{repar},
where it is shown how the somewhat
misterious reparametrization invariance \cite{KoDo} follows from the
permutation symmetry of the $n$ replicas. Ward identities can then be
derived, at least in principle, from the disappearance of the second,
third, etc.\ variations of the free energy funcional.
We have thus two different methods to obtain Ward identities:
invariance under infinitesimal permutations and/or reparametrization.
To make the presentation clear, computational details are left to
the Appendix.

\section{Permutation symmetry and Ward identities}

\label{inf}
Our starting point is the free energy $F$, expressed as a functional of
the order parameters $q_{\alpha\beta}$, $\alpha,\beta =1,2,\ldots,n$ and
$\alpha<\beta$, with $n$ the replica number. The order parameters
$q_{\alpha\beta}$ are usually regarded as the elements of a real,
symmetric matrix, zero along the diagonal 
($q_{\alpha\beta}=q_{\beta\alpha}$, $q_{\alpha\alpha}=0$); for our
purposes it will be useful to think of them as components of a
$\frac{1}{2}n(n-1)$ dimensional {\em vector} $\vec q$. The association
between the matrix $q_{\alpha\beta}$ and the vector $\vec q$ is obvious:
one lists the elements of $q_{\alpha\beta}$ with $\alpha<\beta$ in any
prescribed order (say row by row) to form a column vector.

Although, the symmetry arguments we are going
to use are completely general, we
wish to give here some
specific examples. The reader may think of $F(\vec q)$ as
\begin{itemize}
\item the functional appearing in the integral representation of the
quenched averaged free energy of the SK model \cite{Review};
\item the Legendre-transformed free energy,
with respect to a source $h_{\alpha\beta}$,
of a replica field theory with a
Lagrangean $\cal L(\phi _{\alpha\beta})$ invariant under any permutation
of the $n$ replicas;
\item a functional obtained by a {\em second} Legendre-transformation
\cite{second} of a replica field theory with a Lagrangean $\cal L
(\phi_{\alpha})$ (the source $h_{\alpha\beta}$ now couples to products
like $\phi_{\alpha}\phi_{\beta}$). In this case the notation
$G_{\alpha\beta}$ is preferred to $q_{\alpha\beta}$. The random field
Ising model \cite{RFIM} and the random manifold problem
\cite{manifold,MePa} are good examples where such a functional has an
important role. Although the diagonal elements
$G_{\alpha\alpha}$ are no longer
zero now, this leads to only slight modifications in the arguments and has
no influence on the results.
\end{itemize}

By construction, the free energy is invariant with respect to the
permutations $P$ of the replicas:
\[
   F({\vec q}\,')=F(\vec q) \qquad {\rm for} \qquad q'_{\alpha\beta}=
   q_{P_{\alpha}P_{\beta}} \quad.
\]
Permuting the components of a vector will not change its length, so
$P$ generates an orthogonal transformation:
\[
  \vec {q}\,'={\bf O}\,
   \vec q \quad, \qquad 
   {\bf O}^{\rm T}={\bf O}
   ^{-1} \quad.
\]
The invariance of $F$ under ${\bf O}$
implies that its gradient transforms as
a vector:
\begin{equation}
       \frac{\partial F}{\partial \vec {q}\,'}=
       {\bf O}\,
       \frac{\partial F}{\partial \vec q} \quad. \label{grad}
\end{equation}

Throughout this paper, Eq.~\ref{grad} will be used for a vector 
$q_{\alpha\beta}$ built up by Parisi's hierarchical construction
\cite{Pa,Review}, and having the following properties: The (for the
time being) discrete values of the matrix elements will be called
$q_r$, and the sizes of the hierarchically arranged blocks $p_r$,
$r=0,1,\ldots,R+1$
and, by convention, $p_0=n$ and $p_{R+1}=1$. $r$ is the overlap of
the replicas $\alpha$ and $\beta$, $r=\alpha\cap\beta$, i.e.\ $
q_{\alpha\beta}=q_r$, and by definition $\alpha\cap\alpha=R+1$.
Both series of parameters are assumed to be monotonic: $q_s<
q_r$ and $p_s<p_r$ for $s<r$. $R$ is the number of replica symmetry
breaking steps; our main concern will be the evolution of the
symmetries of the system as $R\to\infty$. For large but finite $R$
the $q$'s and the $p$'s fill the intervals $[q_0,q_R]$ with
$q_0\geq 0$, $q_R<1$, and $[p_1,p_R]$ with $p_1>p_0=n\geq 0$,
$p_R<1$, respectively, in a quasi continuous manner:
\begin{eqnarray}
q_{r+1}-q_r & = & O(1/R)\quad,\qquad \,\,r=0,1,\ldots,R-1
\quad, \nonumber\\
q_{r+1}+q_{r-1}-2q_r & = & O(1/R^2)\quad, \qquad
r=1,2,\ldots,R-1\quad, \nonumber\\
p_{r+1}-p_r & = & O(1/R)\quad, \qquad \,\,r=1,2,\ldots,R-1
\quad, \nonumber \\
p_{r+1}+p_{r-1}-2p_r & = & O(1/R^2)\quad,
\qquad r=2,3,\ldots,R-1 \label{cont}
\end{eqnarray}
(the difference $p_1-p_0$ and $p_{R+1}-p_R$ may be of
$O(1)$). A vector $\vec q$ associated with a matrix
$q_{\alpha\beta}$ with the above structure 
and continuity properties as in Eq.~\ref{cont} will,
in the following,
be called a
Parisi vector.
 Note that we have not assumed the
$q$'s and $p$'s to be stationary. In fact, in most of what
follows we will be considering symmetries that are present
irrespective of whether we are at a stationary point or not.

Let us consider now the action of a special permutation of
replicas on a Parisi vector. The permutation will be chosen in
such a way as to interchange two blocks of size $p_{r+2}$ and
leave the rest unchanged. The replicas belonging to these two
blocks will be labelled by $\alpha_i$ and $\beta_i$,
respectively, $i=1,2,\ldots,p_{r+2}$. The permutation in question
will then act as $P_{\alpha_i}=\beta_i$ and
$P_{\beta_i}=\alpha_i$ for $i=1,2,\ldots,p_{r+2}$, and as
$P_{\alpha}=\alpha$ for $\alpha$ outside the two selected blocks.
If the two blocks belong to the same block of size $p_{r+1}$,
i.e.\ if $\alpha_i\cap\beta_i=r+1$, this permutation is just an
element of the residual symmetry group that remains after replica
symmetry breaking (RSB), and will leave the Parisi vector
invariant ($\vec {q}\,'=\vec q$). If the two blocks are chosen
farther apart ($\alpha_i\cap\beta_i<r+1$), however, $\vec {q}\,'$
will not be a Parisi vector any more, and, depending on $P$,
a smaller or larger difference $\vec {q}\,'-\vec q$ develops.
The smallest change is expected for $\alpha_i\cap\beta_i=r$.
For this choice, the only
nonzero components of $\vec {q}\,'-\vec q$ are the
following:
\[(\vec {q}\,'-\vec q)_{\alpha_i\gamma}=(\vec {q}\,'-\vec q)
_{\beta_i\delta}=q_{r+1}-q_r \quad,\]
for
\[ \alpha_i\cap\gamma=\beta_i\cap\delta=r \quad {\rm and}\quad
 \alpha_i\cap\delta=
\beta_i\cap\gamma=
r+1 \quad,\]
and
\[(\vec {q}\,'-\vec q)_{\alpha_i\gamma}=(\vec {q}\,'-\vec q)
_{\beta_i\delta}=q_r-q_{r+1}\quad,\]
for
\[ \alpha_i\cap\gamma=\beta_i\cap\delta=r+1 \quad {\rm and}\quad
\alpha_i\cap\delta=\beta_i\cap\gamma=r\quad.\]
The length of $\vec {q}\,'-\vec q$ is
\begin{equation}
\sqrt{(\vec {q}\,\,'-\vec q)^2}=\sqrt{\sum_{\alpha<\beta}
(\vec {q}\,\,'-\vec q)^2_{\alpha\beta}}=\sqrt{4p_{r+2}
(p_{r+1}-p_{r+2})(q_{r+1}-q_r)^2}
\label{norm}
\end{equation}
which is of $O(1/R^{3/2})$, i.e.\ infinitesimal.

The idea of constructing an "infinitesimal" permutation has been
around for a long time in the replica approach to random systems. The
first such transformation appeared in \cite{KoNe}, where it was
shown that a suitably chosen linear combination of replicon
eigenvectors added to a Parisi vector will result in a
"reparametrization" of this vector, i.e.\ in a modified sequence of
parameters $q_r$ and $p_r$. This observation will be further developed
below. Shortly thereafter Goltsev introduced a set of "infinitesimal
permutations" and showed how any finite permutation can be built up
from infinitesimal ones \cite{Go2}, without exploiting the full
potential of these infinitesimal generators. Independently, Parisi
and Slanina rediscovered the same transformation in a random polymer
context \cite{PaSla}. Finally an infinitesimal transformation
closely related to the one above was used by ourselves to derive Ward
identities for the spin glass \cite{Cyrano}, in the rest of this section
we recall some of the results derived there\footnote{In Ref.~\cite{Cyrano},
at editing, the equation numbering was messed up. The puzzled reader will
find the correct numbering in the original cond-mat/9802166 version.}.

After finding infinitesimal symmetry transformations, we can follow the
usual steps to obtain Ward identities in a system with a continuous
symmetry \cite{Ward}. Since $\vec {q}\,'$ is very close to $\vec q$ we
can expand the left hand side of Eq.~\ref{grad} to get
\begin{equation}
{\bf O}\, \vec f -\vec f= {\bf M}\,
(\vec {q}\,'-\vec q) +\ldots \label{expand}
\end{equation}
where $\vec f$ is the gradient vector
\begin{equation}
\vec f=\frac{\partial F}{\partial \vec q} \label{f}
\end{equation}
and ${\bf M}$ is the mass operator with components
\begin{equation}
M_{\alpha\beta,\gamma\delta}=\frac{\partial^2 F}{\partial
q_{\alpha\beta} \partial q_{\gamma\delta}} \quad. \label{M}
\end{equation}
Evaluating the derivatives in Eqs.~\ref{f} and \ref{M} at
$q_{\alpha\beta}$ (a Parisi vector) we get a gradient vector
$\vec f$ with the same Parisi-like structure, and a mass operator
which has the structure of an ultrametric matrix. (Ultrametric
matrices commute, by definition, with all the elements of the
residual group. Their structure was studied in detail in
\cite{ultrametric}.) Eq.~\ref{expand} can then be analysed by the
block diagonalization procedure described in \cite{ultrametric}.
As shown in that paper, the relevant quantities representing the
diagonal blocks are:
\begin{itemize}
\item
the one-dimensional replicon (R) blocks (i.e.\ eigenvalues)
$\lambda (r;k,l)$, $r=0,1,\dots,R$; $k,l=r+1,r+2,\ldots,R+1$,
and
\item
the $(R+1)\times (R+1)$-dimensional longitudinal-anomalous (LA)
"kernels" of the LA blocks, $K_k(r,s)$, $r,s=0,1,\ldots,R$
and $k=0,1,\ldots,R+1$.
\end{itemize}
The explicit expressions for $\lambda (r;k,l)$ and $K(r,s)$ are
given in \cite{ultrametric}. After block diagonalization the
various components of Eq.~\ref{expand} give, to leading order in
$1/R$, the following set of equations:
\begin{equation}
f_{r+1}-f_r=(q_{r+1}-q_r) \lambda
(r+1;r+2,r+2)+O(1/R^2)\quad,\label{basic}
\end{equation}
\begin{eqnarray}
\lambda(r+1;r+2,r+2)-\lambda(r;r+1,r+2)&=&O(1/R)\quad,
\nonumber \\
\lambda(r;r+2,r+2)-\lambda(r;r+1,r+2)&=&O(1/R)\quad,
\nonumber\\
K_{r+2}(r,s)-K_{r+2}(r+1,s)&=&O(1/R)\quad
,\label{folyt}
\end{eqnarray}
with $r=0,1,\ldots,R-1$ and $s=0,1,\ldots,R$.
Eq.~\ref{folyt} expresses a continuity property of the mass
operator, while Eq.~\ref{basic} is a Ward identity which
establishes a relationship between the first and second
derivatives of $F$. In the limit $R\to\infty$,
after introducing the continuous parameter $x$ by
\begin{equation}\label{x}
x=\frac{r}{R+1}\quad, \qquad 0\leq x \leq 1 \quad,
\end{equation}
(\ref{basic}) becomes
\begin{equation}
\frac{df(x)}{dx}=\frac{dq(x)}{dx} \, \lambda(x;x,x)\quad, \qquad 0<x<1\quad.
\label{derx}
\end{equation}
If, finally,  we take Eq.~\ref{derx} at a stationary point where $f\equiv 0$,
we obtain
\begin{equation}
\lambda(x;x,x)=0 \label{Goldstone}
\end{equation}
for all $x$ where $\frac{dq(x)}{dx}\neq 0$.
(If, as is often the case, there is
a breakpoint $x_1$ beyond which $q(x)$ and also $\lambda(x;x,x)$ are constant,
(\ref{Goldstone}) still holds in the limit $x\to x_1^{-}$, and, by continuity, it also
holds for $x>x_1$.)
The status of Eq.~\ref{Goldstone} is that of a Goldstone theorem for spin glasses:
under the assumptions of permutation invariance and the existence of the continuous
limit $R\to \infty$, (\ref{Goldstone}) follows, independently of the concrete
form of the free energy functional $F$.

It is straightforward, at least in principle,
to apply the above ideas for deriving higher order Ward identities.
The mass operator $\bf M$ is the second derivative of the free energy, and
thus it transforms as a tensor under an orthogonal transformation
$\bf O$:
\begin{equation}
{\bf M}'=
{\bf O M O^{\rm T}} \label{tensor}\quad,
\end{equation}
where $\bf M'$ refers to the derivatives in (\ref{M}) evaluated at
$\vec{q}\,'={\bf O}\vec{q}$. Expanding the left hand side of
Eq.~\ref{tensor} around $\vec {q}$, and introducing the 3-point vertex
function as
\begin{equation}
W_{\alpha\beta,\gamma\delta,\mu\nu}=\frac{\partial^3 F}{\partial
q_{\alpha\beta}\partial q_{\gamma\delta}\partial q_{\mu\nu}}\quad,
\label{W}
\end{equation}
we get
\begin{equation}
({\bf O M O^{\rm T}})_{\alpha\beta,\gamma\delta}-
({\bf M})_{\alpha\beta,\gamma\delta}= \sum_{\mu<\nu}
W_{\alpha\beta,\gamma\delta,\mu\nu}\, (\vec {q}\,'-\vec q)_{\mu\nu}
+\ldots \qquad. \label{expandW}
\end{equation}
Eq.~\ref{expandW} is the starting point for obtaining the set of
Ward-identities establishing the exact connections between components
of the 2- and 3-point vertices.
To accomplish this work, a procedure, similar to that of block
diagonalisation of the 2-point vertices, transforming the 3-point
vertices into a "canonical" form is needed.
A simple example of such a relationship between 2 and 3-point
vertices is worked out in \cite{Cyrano}. 

\section{Reparametrization invariance -a subtle corollary of permutation
symmetry}

\label{repar}
As we see now, the symmetry responsible for the Goldstone modes is permutation invariance that becomes "continuous" in the limit $R\to\infty$. Nevertheless,
in Ref.~\cite{KoDo} two of us purported to derive (\ref{Goldstone}) from a
rather liberal interpretation of an other kind of symmetry:
reparametrization invariance. The obviously correct result was obtained
by using some uncontrollable calculational steps, such as
forgetting about the
differences between derivatives with respect to $x$ or $p(x)$.
Having two independent symmetries leading to the same exact property of
the system, i.e.\ the masslessness of some of the modes, is rather unlikely,
and one can suspect that a connection between them must exist.
In this section, we will find this connection showing how reparametrization
invariance follows,
in the limit $n\to 0$ and $R\to \infty$,
from the primary symmetry of the equivalence of the $n$
replicas.

\begin{figure} 
\centering\epsfig{figure=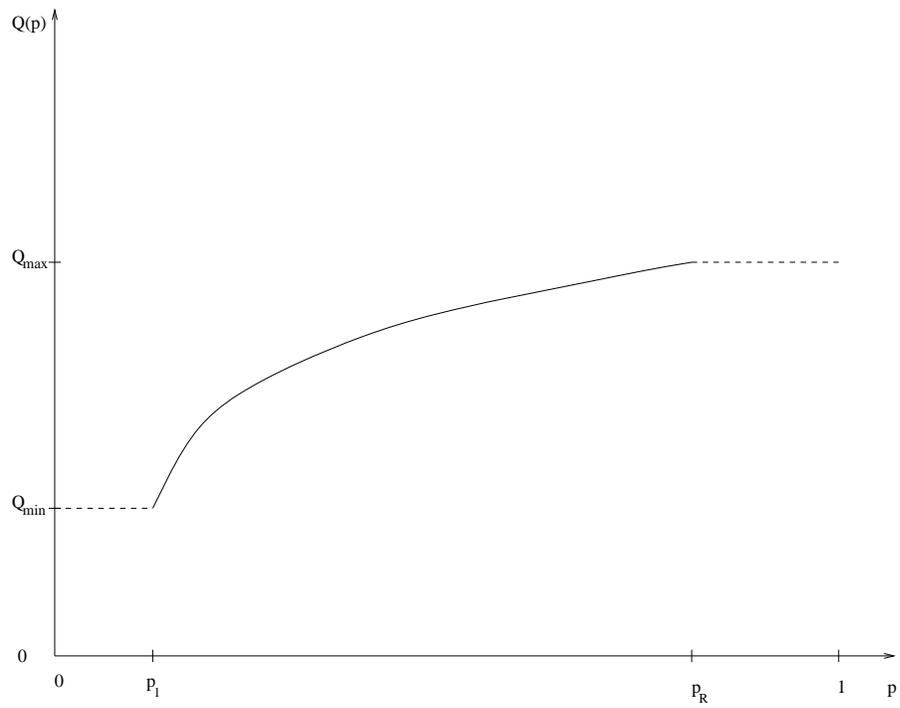,height=.5\textheight,angle=-90}
\caption{$Q(p)$ defines a chain of reparametrization
transformations}
\label{fig1}
\end{figure}

The first thing we need is a clear definition of the reparametrization
{\em transformation}. It connects two Parisi vectors $\vec q\, \to \,
\vec q\,'$ with the following properties:
\begin{eqnarray}
p'_r&=&p_r+\delta p_r \quad, \qquad r=1,\ldots,R \quad;\label{rt1}\\
q'_r&=&Q(p'_r) \quad, \quad \qquad r=1,\ldots,R \quad;\label{rt2}\\
q'_0&=&q_0 \quad,\label{rt3}
\end{eqnarray}
supplemented by the condition that the endpoints be
invariant, $\delta p_1=
\delta p_R=0$. The function $Q(p)$ is obtained from
$q(x)$ and $p(x)$ by eliminating $x$ between them, and it remains fixed
under a chain of reparametrization transformations (see Fig.\ 
\ref
{fig1}).

At this point we wish to stress that the reparametrization invariance
is not a property of the free energy functional $F(q_{\alpha\beta})$,
but of the functional $\tilde F(q_r,p_r)$ obtained from $F$ by restricting
its argument $q_{\alpha\beta}$ to the subspace spanned by the Parisi-like
vectors. This subspace may be called longitudinal, and, by extension,
$\tilde F$ may be called the "longitudinal" free energy. $\tilde F$ remains
unchanged when moving from $\vec q$ to $\vec q\,'$ by the reparametrization
transformation of Eqs.~\ref{rt1}-\ref{rt3}:
\begin{equation}
\tilde F(q_r',p_r')=\tilde F(q_r+\delta q_r,p_r+\delta p_r)=
\tilde F(q_r,p_r)\quad.
\label{reparinv}
\end{equation}
In the above equation $\vec q$ and $\vec q\,'$ are Parisi vectors with the
properties given in Eq.~\ref{cont}, it expresses, therefore, an exact symmetry
of $\tilde F$ only in the limit $R\to \infty$. For finite $R$ the invariance
of $\tilde F$ is only approximate, valid within corrections of $O(1/R)$.
Eq.~\ref{reparinv} can be cast into an equivalent form using variations
of $\tilde F$:
\begin{equation}\label{variations}
\delta^{(k)}\tilde F[q(x),p(x)]=0
\end{equation}
for any positive integer $k$. In Eq.~\ref{variations} only $p(x)$ is
varied independently, with the constraint of fixed endpoints
$\delta p(0)=\delta p(1)=0$. By Eq.~\ref{rt2}, $\delta q(x)$ can be
expressed in terms of $\delta p(x)$, an expanded form of which is useful
to be recorded here for further reference:
\begin{equation}\label{deltaq}
\delta q(x)=Q'[p(x)] \delta p(x)+\frac{1}{2} Q''[p(x)] \delta p(x)^2+
\ldots\qquad.
\end{equation}

To prove the invariance properties in Eqs.~\ref{reparinv} or
\ref{variations}, $\Delta \tilde F\equiv \tilde F(q'_r,p'_r)-
\tilde F(q_r,p_r)$ will be expanded up to second order in the
$\delta q_r$'s and $\delta p_r$'s. A simple Taylor expansion yields
\begin{equation}\label{Taylor}
\Delta \tilde F=\sum_{r=1}^{R}\left(
\frac{\partial \tilde F}{\partial
 q_r}\delta q_r + 
\frac{\partial \tilde F}{\partial p_r}
\delta p_r \right) +\frac{1}{2}\sum_{r,s=1}^{R}\left(
\frac{\partial^2 \tilde F}{\partial q_r\partial q_s}\delta q_r
\delta q_s+
2\frac{\partial^2 \tilde F}{\partial q_r\partial p_s}
\delta q_r\delta p_s+
\frac{\partial^2 \tilde F}{\partial p_r\partial p_s}
\delta p_r\delta p_s\right).
\end{equation}
On the other hand, the displacement between the two
"longitudinal" (Parisi-type) vectors in the full
vector space of the $q_{\alpha\beta}$'s gives an equivalent
form for $\Delta \tilde F$. Using the short-hand notation of
brackets as in the Appendix (Eqs.~\ref{bracketf},\ref{bracketM}),
we can write
\begin{equation}\label{fullTaylor}
\Delta \tilde F=
 \langle f \,|\, q'-q \rangle+\frac{1}{2}
 \langle q'-q \,|\, M \,|\,q'-q \rangle+\ldots\quad.
\end{equation}
$\vec f$ and $\bf M$ were defined in Eqs.~\ref{f} and \ref{M},
and under {\em any} permutation $P$ of the replicas, corresponding
to an orthogonal transformation $\bf O$ in the order parameter
space of the $q_{\alpha,\beta}$'s, they transform according to
Eqs.~\ref{grad} and \ref{tensor}, respectively. As
pointed out in Sec.~\ref{inf}, this is a clear consequence of the
permutation invariance of the free energy $F(q_{\alpha\beta})$.
Furthermore, taking the permutation $P$ from the residual symmetry
group corresponding to the ultrametric construction of the vector
$\vec q$ (which is obviously different from the group of transformations
defined by $\vec q\,'$),
i.e.\ ${\bf O}\vec q=\vec q$, we get
\begin{equation}\label{ultrametric}
\vec f\,'=\vec f={\bf O}\vec f\quad 
{\rm and}\quad
{\bf M}'{\bf O}=  
{\bf M}{\bf O}={\bf O}{\bf M}.
\end{equation}
Thus $\vec f$ and ${\bf M}$ are ultrametric, and Eqs.~\ref{fq},\ref
{qMq} of the Appendix can be used in Eq.~\ref{fullTaylor}.  
Putting everything together from the Appendix (especially
from Eqs.~\ref{KR}, \ref{KL}, \ref{kernel} and \ref{scalarproduct}),
a comparison of Eqs.~\ref{Taylor} and \ref{fullTaylor} yields:
\begin{eqnarray}
\frac{\partial\tilde F}{\partial q_r}
&=&\frac{n}{2}(p_r-p_{r+1})f_r
\quad\qquad\qquad\qquad\qquad\qquad\qquad\qquad\qquad
 r=0,\dots,R\quad, \nonumber\\
\frac{\partial\tilde F}{\partial p_r}
&=&\frac{n}{2}
(q_r-q_{r-1})
\left[f_r-\frac{1}{2}
(q_r-q_{r-1})\lambda(r;r+1,r+1)\right]\qquad r=1,\ldots,R\quad.
\label{partialp}
\end{eqnarray} 
From this,
in the limit $R\to \infty$, 
the first variation $
\delta^{(1)}\tilde F
$ follows
immediately.
Introducing the continuous variable $x$ (see Eq.~\ref
{x}):
\begin{equation}\label{firstvar}
\delta^{(1)}\tilde F=\frac{n}{2}\int_0^1dx\left[q'(x)f(x)-
Q'[p(x)]p'(x)f(x)\right]\,\delta p(x)\quad;
\end{equation}
in
deriving the above expression,
use has been made
of Eqs.~\ref{Taylor} and \ref{partialp},
together with the reparametrization
constraint Eq.~\ref{deltaq}.

The integrand in Eq.~\ref{firstvar} is identically zero,
which is a direct consequence of the definition of
$Q(p)$: $Q[p(x)]=q(x)$. Thus the disappearence of the first
variation, $\delta^{(1)}\tilde F=0$, seems to be somewhat trivial.
We must emphasize, however, that it is true along the whole
chain of reparametrization transformations; a result which follows
from the crucial step in the proof: if $F(q_{\alpha\beta})$ is
invariant for any permutations of the replicas, then its derivatives
taken at an ultrametrically structured $\vec q$ are themselves
ultrametric. The permutation group of the $n$ replicas is
"large" enough to include all the subgroups defined by the
Parisi-type vectors along a path of consecutive infinitesimal
reparametrization transformations. $\delta^{(1)}\tilde F$ is,
therefore, {\em identically} equal to zero, implying the two
equivalent form of reparametrization invariance, Eqs.~\ref
{reparinv} and \ref{variations}.

With its proof now accomplished,
we can proceed and use Eq.~\ref{variations} for $k=2,3,\ldots$
to derive Ward-identities (in principle to any desired order).
They must not be different from those following from the
primary symmetry of permutation invariance using infinitesimal
permutations, as in Sec.~\ref{inf}. To show this, we compute
$\delta^{(2)}\tilde F$. The second partial derivatives of
$\tilde F$ can be calculated, by intensive use of the Appendix,
just like the first ones were in Eq.~\ref{partialp}. 
The results are as follows:
\begin{eqnarray}
\frac{\partial ^2\tilde F}{\partial q_r\partial q_s}&=&
\frac{n}{4}
(p_r-p_{r+1})
(p_s-p_{s+1})K_0(r,s)+
\frac{n}{2}
(p_r-p_{r+1})\lambda(r;r+1,r+1)\,
\delta_{r,s}^{\rm Kr}
\nonumber\\
&&r,s=0,\dots,R\quad,\nonumber\\
\frac{\partial ^2\tilde F}{\partial p_r\partial q_s}&=&
\frac{n}{4}
(q_r-q_{r-1})
(p_s-p_{s+1})K_0(r,s)+
\frac{n}{2}f_r\,
\delta_{r,s}^{\rm Kr}+\nonumber\\
&&\mbox{}+\frac{n}{2} \left[-f_r+
(q_r-q_{r-1})\lambda(r;r+1,r+1)\right]\,
\delta_{r-1,s}^{\rm Kr}\label{partialpp}\\
&&r=1,\ldots,R\quad{\rm and}\quad s=0,\ldots,R\quad,\nonumber\\
\frac{\partial ^2\tilde F}{\partial p_r\partial p_s}&=&
\frac{n}{4}
(q_r-q_{r-1})
(q_s-q_{s-1})K_0(r,s)\nonumber\\
&&r,s=1,\dots,R\quad.\nonumber
\end{eqnarray}
It is now straightforward to compute $\delta^{(2)}\tilde F$
using Eqs.~\ref{deltaq},\ref{Taylor},\ref{partialp} and
\ref{partialpp}. Note that the second term in Eq.~\ref{deltaq}
multiplied by the first partial derivative of $\tilde F$ with
respect to $q(x)$ also contributes. The rather lengthy
calculation consists of mainly partial integrations, surface
terms always disappearing because of the constraint $\delta
p(0)=\delta p(1)=0$. After a lot of simplification, the
$K_0(x,y)$ terms cancel each other, yielding the surprisingly
simple result:
\begin{equation}\label{secondvar}
\delta^{(2)}\tilde F=\frac{n}{2}\int_0^1dx
\,\left[q'(x)\lambda(x;x,x)-f'(x)\right]\,
Q'[p(x)]\,\delta p(x)^2\quad.
\end{equation}
Since $\delta p(x)$ is arbitrary, and assuming a monotonic $Q(p)$,
the vanishing of the second variation is equivalent with the
vanishing of the expression in the brackets in Eq.~\ref{secondvar};
i.e.\ Eq.~\ref{derx} is now regained from the invariance of the
free energy $\tilde F$ when the "gauge" $p(x)$ is changing.

\begin{figure}
\centering
\epsfig{figure=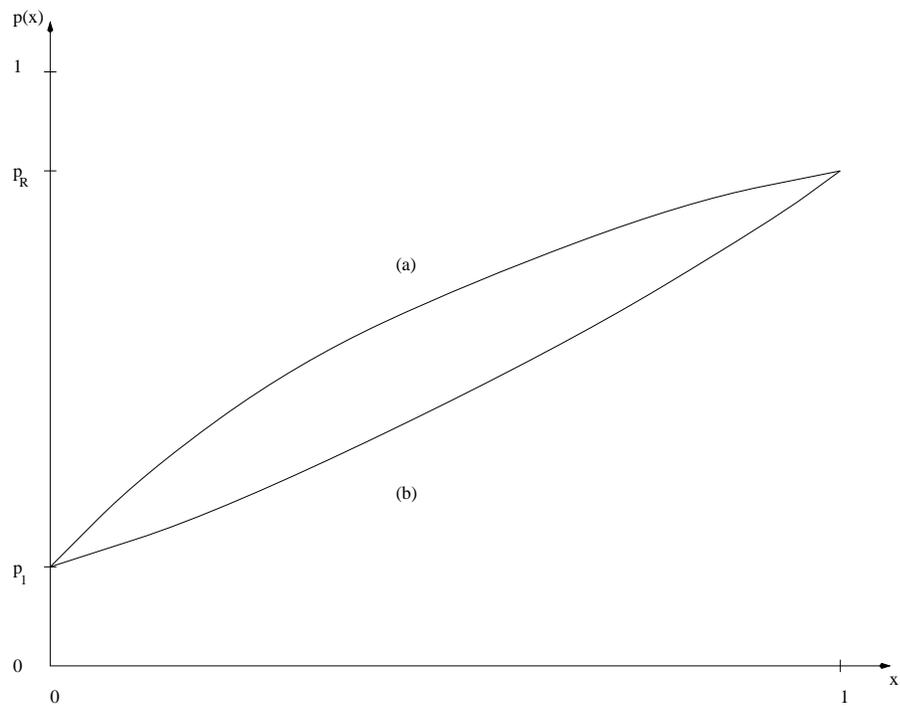,height=.5\textheight,
angle=-90}
\caption{Reparametrization with fixed endpoints}\label{fig2}
\end{figure}

\begin{figure}
\centering
\epsfig{figure=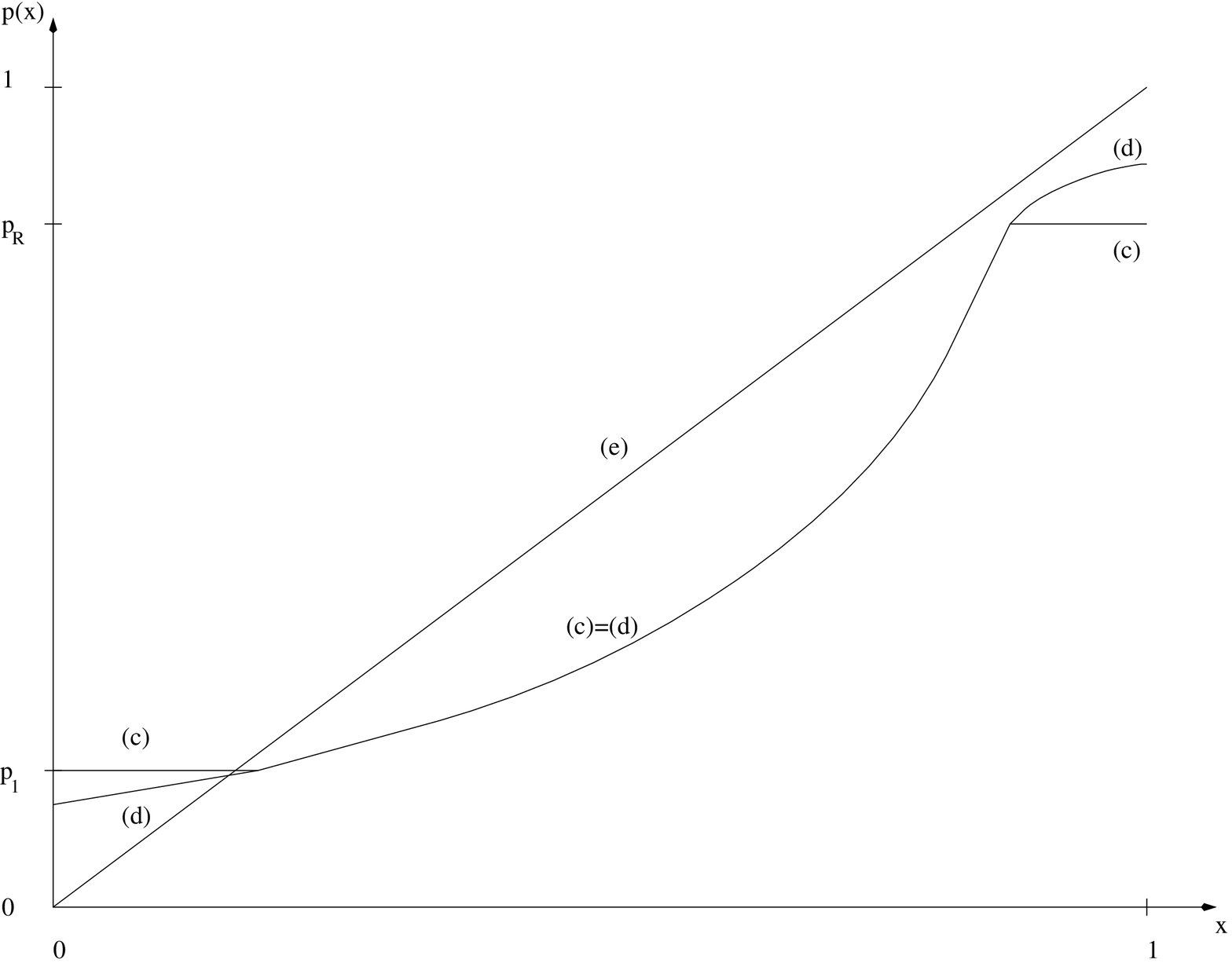,height=.5\textheight,
angle=-90}
\caption{Extending reparametrization transformation to gauges
with moving endpoints (see text)}\label{fig3}
\end{figure}

An example of two equivalent "gauges" with fixed endpoints is shown
in Fig.~\ref{fig2}. We can even deform curve (b) to curve (c) in
Fig.~\ref{fig3} with the plateaux around $x=0$ and $x=1$. Supplementing
the definition of $Q(p)$ by the plateaux regions drawn by dashed lines
in Fig.~\ref{fig1}, we can easily figure out that the parametrization
of curve (d) represents the {\em same} Parisi vector $q_{\alpha\beta}$
as curve (c). In that case, $q(x)$ has also two plateaux with
$Q_{\rm min}$ and $Q_{\rm max}$. Thus, the "longitudinal" free
energy remains invariant for reparametrizations where the endpoints
$p(0)$ and $p(1)$ move apart: $0\le p(0)\le p_1$ and
$p_R\le p(1)\le 1$. The most common gauge,
$p(x)=x$, introduced by Parisi \cite{Pa}, is also displayed
as curve (e) in Fig.~\ref{fig3}.

As a last remark, we want to stress that reparametrization
invariance expresses simply the fact that it is
not the gauge $p(x)$ but $Q(p)$ which
has a physical meaning. (For the Ising spin glass, the
probability distribution of pures states is the derivative
of the inverse function $p(Q)$.)

\section*{Acknowledgment}

This work has been supported by the Hungarian Science Fund
(OTKA), Grant No.~T032424, and by the convention CEA/MAE 2000.

\section*{Appendix}

In the starting phase of Parisi's ultrametric construction (see the
references \cite{Pa,Review}) the block sizes $p_r$'s are large integers
with the property that each $p_{r+1}$ is a divisor of $p_r$ for all
$r=0,1,\ldots,R$ and the continuum of the $p_r$'s with the inverted
monotonicity of a Parisi vector, defined in Sec.~\ref{inf}, is obtained
after taking $n\to 0$, $R\to \infty$. (For a more mathematical
treatment, see \cite{padic}.) We adopt a similar construction in
the reparametrization procedure and consider $p_{r+1}$ a common
divisor of $p'_r$ and $p_r$, while $p'_r$ itself is a divisor of
$p_r$; of course, this is valid at the stage before the "upside down"
continuation of $n$ from a large integer to zero, and $\delta p$ of
Eq.~\ref{rt1} can be considered an arbitrary function of $x$ with the
only condition $\delta p(0)=\delta p(1)=0$. 
 
In this appendix, however, we remain in the domain of large integer
block sizes
$p_r$ and $p_r'$, for all $r=0,1,\ldots,R$ and finite $R$, and the
properties of the previous paragraph will be assumed. This enables us
to expand the
vector $\vec{q}\,'-\vec q$, the displacement
under the reparametrization transformation, in terms of
basis vectors. These basis vectors form a complete set in the
$\frac{n(n-1)}{2}$-dimensional space of the $q_{\alpha\beta}$'s,
and they block-diagonalize any generic ultrametric matrix. (For a
detailed analysis of the structure of this non-orthogonal basis,
see \cite{ultrametric}.) Following \cite{KoNe}, we can figure out
that only two types of basis vectors are involved in the expansion:
a replicon one $
\vec q\,^{({\rm R},r)}
$ and a longitudinal one
$\vec q\,^{({\rm L},r)}$.
$\vec q\,^{({\rm R},r)}$ is just the linear
combination of the $(r;r+1,r+1)$ modes
which shows up in Eq.~(7) of \cite{KoNe}, and as
such, it is an eigenvector of any ultrametric matrix with the
eigenvalue $\lambda (r;r+1,r+1)$. It has nonzero components only for
$\alpha\cap\beta=r$, taking two different values depending on the
overlap of $\alpha$ and $\beta$ with respect to the ultrametric
structure defined by the new block sizes $p_r'$-s:
\begin{eqnarray} 
\vec q\,^{({\rm R},r)}
_{\alpha\beta}&=& \frac{(p_r-2p_{r+1})
(p_r-p_r')}{2 p_{r+1}^2},\qquad\quad\!\!
 \mbox{$\alpha,\beta$ in
the same $p_r'$ block,} \nonumber \\
\vec q\,^{({\rm R},r)}_{\alpha\beta}&=& -\frac{(p_r-2p_{r+1})
(p_r'-p_{r+1})}{2p_{r+1}^2},\quad\! \mbox{$\alpha,\beta$ in different
$p_r'$ blocks.} \label{repr} 
\end{eqnarray}
As for the longitudinal vector $\vec q\,^{({\rm L},r)}$, it has zero
elements everywhere, except for $\alpha\cap\beta=r$:
\begin{eqnarray}
\vec q\,^{({\rm L},r)}_{\alpha\beta}&=&1,\qquad \alpha\cap\beta=r\,;
\nonumber\\
\vec q\,^{({\rm L},r)}_{\alpha\beta}&=&0,\qquad \alpha\cap\beta\not=r\,.
\label{longr}\end{eqnarray}
$r=0,1,\ldots,R$ in Eq.~\ref{longr}, while, remembering that
$p_0'=p_0=n$, the replicon vector of
Eq.~\ref{repr} is not defined for $r=0$. 

The displacement vector $
\vec q\,'-\vec q
$ can be expanded as
\begin{equation}\label{expansion}
\vec q\,'-\vec q
=\sum_{r=1}^{R} \left(
K^{({\rm R},r)}
\,\vec
 q\,^{({\rm R},r)}
+K^{({\rm L},r)}
\,\vec q\,^{({\rm L},r)} \right) + (q_0'-q_0)
\,\vec q\,^{({\rm L},0)}\quad,
\end{equation}
with the coefficients 
$K^{({\rm R},r)}$ and
$K^{({\rm L},r)}$ determined from the conditions
\begin{eqnarray*} 
(\vec q\,'-\vec q)_{\alpha\beta}
&=& q'_r-q_r\, \quad\qquad
\mbox{$\alpha,\beta$ in the same $p'_r$ block,}\\
(\vec q\,'-\vec q)_{\alpha\beta}&=& q_{r-1}'-q_r\, \qquad
\mbox{$\alpha,\beta$ in different $p'_r$ blocks,}
\end{eqnarray*}
for $r=\alpha\cap\beta=1,\ldots,R$. (The $r=0$ case is trivial, with 
$(\vec q\,'-\vec q)_{\alpha\beta}=q'_0-q_0$, leading to the simple
last term in Eq.~\ref{expansion}.) Using Eqs.~\ref{repr},\ref{longr},
it is straightforward to obtain
$K^{({\rm R},r)}$ and
$K^{({\rm L},r)}$. For later reference, it is useful to express  
them in terms of $\delta q_r=q'_r-q_r$ and $\delta p_r=
p'_r-p_r$:
\begin{eqnarray}
K^{({\rm R},r)}
 &=& \frac{2p_{r+1}^2}{(p_r-2p_{r+1})(p_r-p_{r+1})}
\, \left[ (q_r-q_{r-1})+\delta q_r -\delta q_{r-1} \right]\,,
\label{KR}\\
K^{({\rm L},r)}
 &=& \delta q_r - \frac{q_{r-1}-q_r}{p_r-p_{r+1}}
\,\delta p_r - \frac{1}{p_r-p_{r+1}}\,(\delta q_{r-1}\delta p_r-
\delta q_r\delta p_r)\,.\label{KL}
\end{eqnarray}

What we need in the main text, is the scalar product
\begin{equation}\label{bracketf}
 \langle f \,|\, q'-q \rangle
\equiv \sum_{\alpha<\beta}
f_{\alpha\beta}\,(q'-q)_{\alpha\beta}
\end{equation} 
and the matrix elements
\begin{equation}\label{bracketM}
 \langle q'-q \,|\, M \,|\,q'-q \rangle
 \equiv \sum_{\alpha<\beta,
\gamma<\delta} (q'-q)_{\alpha\beta}\, M_{\alpha\beta,\gamma\delta}\,
(q'-q)_{\gamma\delta}
\end{equation} 
for an ultrametrically structured vector $\vec f$ and
matrix $\bf M$. Since $\vec f$ is now a longitudinal vector (orthogonal
to any replicon one), and also using Eq.~\ref{longr}, it follows:
\begin{eqnarray}
\langle f\, |  
 \,q^{({\rm R},r)}
\rangle
&=&0\quad,\nonumber\\
\langle f\, |  
 \,q^{({\rm L},r)}\rangle&=& \frac{n}{2}(p_r-p_{r+1})f_r\quad.
\label{fbasic}
\end{eqnarray}
Combining Eqs.~\ref{expansion} and \ref{fbasic},
\begin{equation}
 \langle f\,|\,q'-q\rangle=\frac{n}{2}\sum_{r=1}^R
(p_r-p_{r+1})f_r
K^{({\rm L},r)}+\frac{n}{2}(n-p_1)f_0\,\delta q_0
\label{fq}
\end{equation}
obtains. To find a similar formula for the matrix element, we
must use results from Ref.~\cite{ultrametric}:
\begin{eqnarray}
{\bf M}\,
\vec q\,^{({\rm R},r)}&=&
\lambda(r;r+1,r+1)\,
\vec q\,^{({\rm R},r)}\quad,\nonumber\\
{\bf M}\,
\vec q\,^{({\rm L},r)}&=&\sum_{s=0}^{R}M^{(0)}_{s,r}\,
\vec q\,^{({\rm L},s)}\quad,
\label{Mbasic}
\end{eqnarray}
where the block matrix element $M^{(0)}_{s,r}$ can be expressed by the
longitudinal kernel $K_0(s,r)$ using Eq.~44 of Ref.~\cite{ultrametric}:
\begin{equation}
M^{(0)}_{s,r}=\lambda(r;r+1,r+1)\,\delta_{s,r}^{\rm Kr}+
\frac{1}{2}(p_r-p_{r+1})\,K_0(s,r)\quad.
\label{kernel}
\end{equation}
From Eqs.~\ref{expansion} and \ref{Mbasic}, it is now straightforward
to find the following expression for the matrix elements:
\begin{eqnarray}
\lefteqn{\langle q'-q \,|\, M \,|\,q'-q \rangle=}\nonumber\\
&&\!\!\!\!\!\!\!\!\!\!\sum_{r=1}^{R}
{K^{({\rm R},r)}}
^2\,
\lambda(r;r+1,r+1)\,
\langle 
q^{({\rm R},r)}
\, |  
 \,q^{({\rm R},r)}
\rangle 
+\sum_{r,r'=1}^R
K^{({\rm L},r)}
K^{({\rm L},r')}\,M^{(0)}_{r,r'}\,
\langle 
q^{({\rm L},r)}
\, |  
 \,q^{({\rm L},r)}
\rangle +\nonumber\\ 
&&\!\!\!\!\!\!\!\!\!\! \mbox{}+\delta q_0\,\sum_{r=1}^R
K^{({\rm L},r)}\,\left( M^{(0)}_{r,0}\,
\langle 
q^{({\rm L},r)}
\, |  
 \,q^{({\rm L},r)}\rangle
+M^{(0)}_{0,r}\,
\langle 
q^{({\rm L},0)}
\, |  
 \,q^{({\rm L},0)}\rangle
\right)+\delta q_0^2\,M^{(0)}_{0,0}\,
\langle 
q^{({\rm L},0)}
\, |  
 \,q^{({\rm L},0)}\rangle
. \label{qMq}
\end{eqnarray}
The scalar products occuring in the above formula can be easily
computed from the definitions in Eqs.~\ref{repr} and \ref{longr}:
\begin{eqnarray}
\langle 
q^{({\rm R},r)}
\, |  
 \,q^{({\rm R},r)}
\rangle& =&-\frac{n(p_r-2p_{r+1})^2 (p_r-p_{r+1})}{8p_{r+1}^3}
\left(\frac{ p_r-p_{r+1}}{p_{r+1}}\,\delta p_r+\frac{1}{p_{r+1}}
\,\delta p_r^2\right) \nonumber \\
\langle 
q^{({\rm L},r)}
\, |  
 \,q^{({\rm L},r)}\rangle &=&
\frac{n(p_r-p_{r+1})}{2}\quad.\label{scalarproduct}
\end{eqnarray}

\end{document}